\documentclass[journal]{IEEEtran}
\usepackage{cite}

\ifCLASSINFOpdf
\usepackage[pdftex]{graphicx}
\else
\fi
\usepackage{amsmath}
\usepackage{amsthm}
\newtheorem{definition}{Definition}

\usepackage{hyperref}
\hypersetup{
    colorlinks=true,
    linkcolor=blue,
    filecolor=magenta,
    urlcolor=cyan,
    pdftitle={Sharelatex Example},
    pdfpagemode=FullScreen,
    }

\newcommand*{\wt}{w^{(t)}}
\newcommand*{\Vt}{V^{(t)}}

\usepackage[ruled,linesnumbered]{algorithm2e}

\usepackage{array}
\usepackage{booktabs}
\usepackage{multirow}
\usepackage{siunitx}

\usepackage[caption=false,font=footnotesize]{subfig}

\usepackage{float}
\usepackage{CJK}

\usepackage{color,soul}
\usepackage{xcolor}

\newcommand{\hlc}[2][yellow]{{\sethlcolor{#1}\hl{#2}}}
\definecolor{c1}{RGB}{204,191,240}
\definecolor{c2}{RGB}{193,237,199}
\definecolor{c3}{RGB}{246,217,176}
\definecolor{c4}{RGB}{246,177,179}
\definecolor{c5}{RGB}{173,220,244}
\usepackage{url}

\begin{document}
%

\title{Provably Secure Disambiguating Neural Linguistic Steganography}

%
%

\author{Yuang~Qi, Kejiang~Chen, Kai~Zeng, Weiming~Zhang, and Nenghai~Yu
\IEEEcompsocitemizethanks{\IEEEcompsocthanksitem This work was supported in part by the National Natural Science Foundation of China under Grant 62472398, U2336206, 62402469, and 62121002. 
\IEEEcompsocthanksitem All the authors are with Anhui Province Key Laboratory of Digital Security, School of Cyber Science and Technology, University of Science and Technology of China, Hefei 230026, China.
\IEEEcompsocthanksitem Corresponding author: Kejiang Chen (Email: chenkj@ustc.edu.cn).}
}

\maketitle

\begin{abstract}
Recent research in provably secure neural linguistic steganography has overlooked a crucial aspect: the sender must detokenize stegotexts to avoid raising suspicion from the eavesdropper. The segmentation ambiguity problem, which arises when using language models based on subwords, leads to occasional decoding failures in all neural language steganography implementations based on these models. Current solutions to this issue involve altering the probability distribution of candidate words, rendering them incompatible with provably secure steganography. 
We propose a novel secure disambiguation method named SyncPool, which effectively addresses the segmentation ambiguity problem. We group all tokens with prefix relationships in the candidate pool before the steganographic embedding algorithm runs to eliminate uncertainty among ambiguous tokens. To enable the receiver to synchronize the sampling process of the sender, a shared cryptographically-secure pseudorandom number generator (CSPRNG) is deployed to select a token from the ambiguity pool. SyncPool does not change the size of the candidate pool or the distribution of tokens and thus is applicable to provably secure language steganography methods. We provide theoretical proofs and experimentally demonstrate the applicability of our solution to various languages and models, showing its potential to significantly improve the reliability and security of neural linguistic steganography systems.
\end{abstract}


%
\IEEEpeerreviewmaketitle

\section{Instruction}


\IEEEPARstart{T}{ext} is the most widely used information carrier in daily life~\cite{dai2010text}, making linguistic steganography possess significant research value and practical significance. 
Generative linguistic steganography methods~\cite{fang2017generating,yang2018rnn,ziegler2019neural,yang2020vae,zhou2021linguistic,xiang2022generative} directly transform secret information into innocuous-looking stegotext during the generation process of language models. Its objective is to effectively deceive machine steganalysis through provable means~\cite{ziegler2019neural}.
Decades ago, researchers began the pursuit of steganography techniques that are \textit{provably secure}~\cite{cachin1998information,hopper2002provably,backes2005public}. According to the information-theoretic model for steganography proposed by Cachin~\cite{cachin1998information} in 1998, the security of a steganographic system (stegosystem) can be quantified in terms of the distribution difference between cover and stego, which is measured by Kullback–Leibler divergence (KL divergence) between the cover distributions and stego distributions. Zero KL divergence means that the steganographic system is perfectly secure. From another perspective, Hopper \textit{et al.}~\cite{hopper2002provably} formalized a provably secure system based on computational complexity. Security is defined as a probabilistic
polynomial time (PPT) distinguisher that cannot distinguish between the cover and stego. Namely, the distributions of the cover and stego objects are computationally indistinguishable. A steganography method that can prove that the distributions of cover and stego are consistent or indistinguishable is known to be provably secure. 


In recent years, the increasing popularity of deep generative models~\cite{goodfellow2014generative,arjovsky2017wasserstein,song2019generative,jozefowicz2016exploring,zhao2023survey} has gradually fulfilled the requirements of provably secure steganography for perfect samplers and explicit data distributions, making it feasible for provably secure steganography to be applied in real-world scenarios~\cite{chen2018provably}. 
For textual carriers, several methods~\cite{ziegler2019neural,zhang2021provably,kaptchuk2021meteor,de2022perfectly,dingDiscopProvablySecure2023a,zhang2024provably} based on neural language deep generative models have been proposed in recent years.
These methods represent valuable attempts to achieve provably secure linguistic steganography.
In general, provably secure neural linguistic steganography can be viewed as a process of message-driven sampling using a generative language model. At each generation step, the language model predicts a probability distribution of the next token to be generated. 
A sender selects a token from this probability distribution to add to the stego sequence using a steganography embedding algorithm with message bits and a chosen key while maintaining the probability distribution. 


\begin{figure}[t]
    \centering
    \includegraphics[width = \columnwidth]{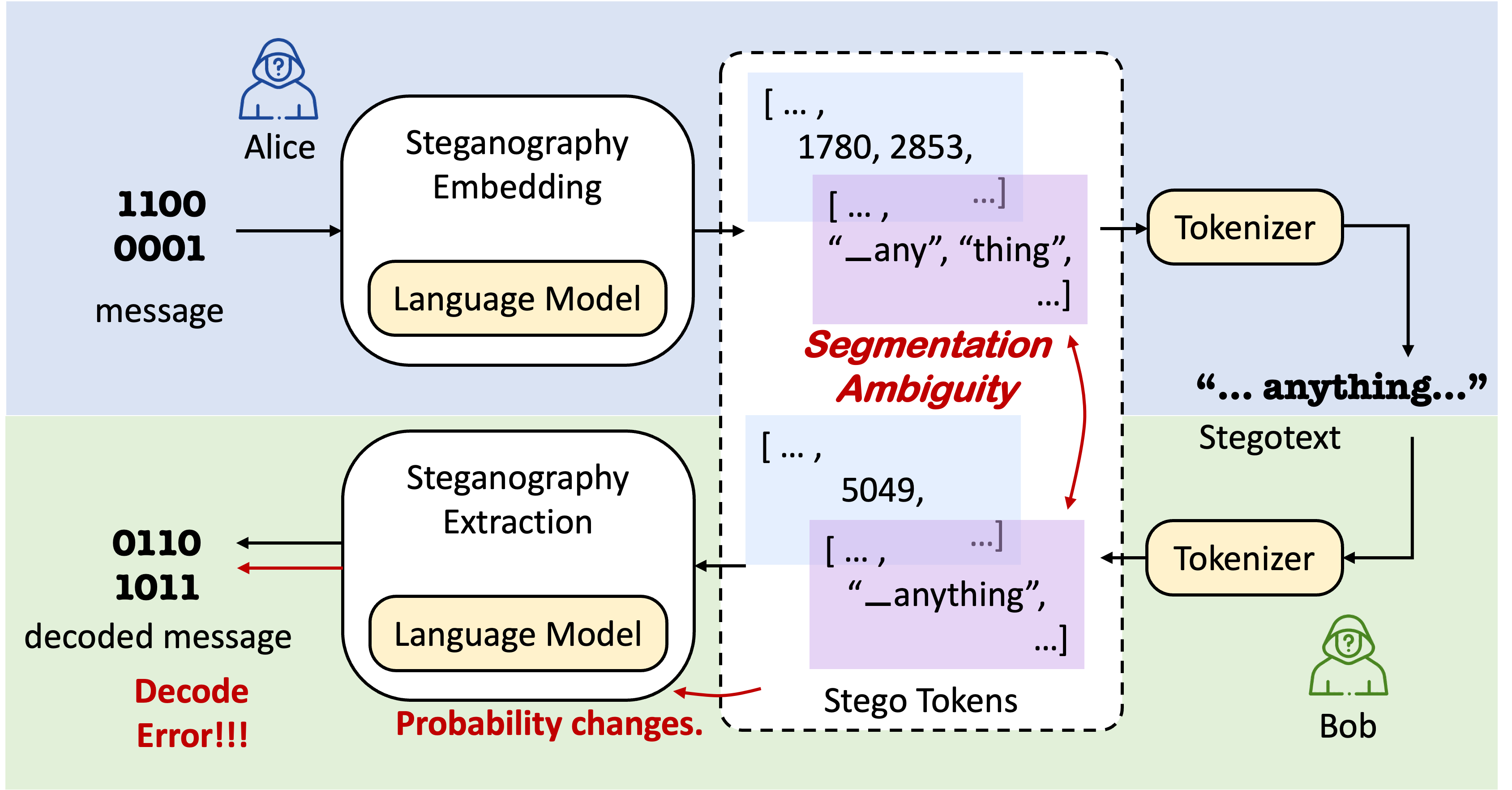}
    \caption{An example of \textbf{segmentation ambiguity} in generative linguistic steganography. The sender Alice generates a token sequence corresponding to subwords ``\_any" and ``thing" during steganography embedding. During transmission, the stego-tokens are decoded into the text ``\ anything". Unfortunately, the receiver Bob may retokenize ``\ anything" as a single token ``\_anything". 
    This can lead to errors in steganography extraction.}
    \label{fig_ambiguity}
\end{figure}

However, the aforementioned neural linguistic steganography methods have either overlooked or intentionally avoided the fact that Alice (steganographic sender) must detokenize the secret tokens into stegotexts before transmitting them to Bob (steganographic receiver) in order to evade Eve's suspicion (steganalyzer). Additionally, the receiver needs to retokenize the received secret text to extract the message. 
Ueoka \textit{et al.}~\cite{ueoka2021frustratingly} were the first to point out that there is no $100\%$ guarantee that Bob could recover the original tokens from detokenized texts, leading to decoding failures. 
The related process is illustrated in \autoref{fig_ambiguity}.
Moreover, since the currently high-performing large language models~\cite{radford2019language,song2019generative,brown2020language} are mostly autoregressive Transformer~\cite{vaswani2017attention} architectures, incorrect segmentation by Bob will lead to changes in the probability distribution corresponding to all subsequent tokens, further affecting message extraction. \textit{Simply adding error-correcting code to messages is not effective.}


This phenomenon is termed \textbf{segmentation ambiguity} by Nozaki and Murawaki~\cite{nozaki2022addressing}.
While segmentation ambiguity has long been a challenging issue for scriptio continua or writing systems without explicit word boundaries, the widespread use of subwords~\cite{sennrich2015neural,kudo2018subword,kudo2018sentencepiece}, which coincided with the invention of the Transformer~\cite{vaswani2017attention} architecture, implies that it now affects any language. 
As long as language models are based on subwords, segmentation ambiguity may arise, thus affecting the normal use of steganography. 
Until recently, the provably secure steganography algorithms proposed by de Witt \textit{et al.}~\cite{de2022perfectly} and Ding \textit{et al.}~\cite{dingDiscopProvablySecure2023a} did not considered the impact of segmentation ambiguity, rendering these steganography methods unusable in recently popular language models.
One approach is to directly transmit tokens to avoid segmentation. However, this behavior is highly questionable in practical steganography applications and contradicts the original intent of steganography.
Therefore, we strongly endorse the call made by Nozaki \textit{et al.}~\cite{nozaki2022addressing} that researchers must consider detokenization and retokenization as \textit{necessary steps} in linguistic steganography.


Several existing methods~\cite{ueoka2021frustratingly,nozaki2022addressing,yan2023secure} have attempted to address the issue of segmentation ambiguity during the steganography process. 
However, we have found that they are not applicable to provably secure neural language steganography. 
Ueoka \textit{et al.}~\cite{ueoka2021frustratingly} attempted to bypass segmentation ambiguity by simply skipping subwords. Their approach, an edit-based steganography with a masked language model, cannot be applied to generative steganography methods. 
Nozaki and Murawaki~\cite{nozaki2022addressing} extended this idea of skipping subwords to generative steganography, where during steganography sampling, tokens whose mapping words are prefixes of others in the candidate pool are removed, ensuring that the receiver cannot find more than one matching token when attempting to extract the message. 
Furthermore, Yan \textit{et al.}~\cite{yan2023secure} aimed to enhance steganography security by maximizing the probability sum of the remaining tokens when removing tokens from the candidate pool. 
However, all of these methods alter the probability distribution of the current token to be generated by the language model. As previously mentioned, provably secure steganography requires maintaining the probability distribution unchanged from normal generation. Therefore, none of the existing solutions for segmentation ambiguity are suitable for provably secure steganography.

To deploy provably secure steganography for practical use, we propose a novel disambiguating method applicable to provably secure steganography, named SyncPool. 
We realize that the fundamental reason for segmentation ambiguity is the loss of information entropy contained in tokens with prefix relationships during detokenization and retokenization. 
In this paper, we group all candidate tokens whose mapping words have a common prefix into several ambiguity pools during generation, which could elimate redundant information entropy and prevent the loss of information. However, the sender and receiver still can not synchronize the token selected in the ambiguity pool.
Therefore, we develop synchronized sampling based on a shared cryptographically secure pseudo-random number generator (CSPRNG) to select a token from the ambiguity pool.
This disambiguation method based on ambiguity pool grouping and synchronized sampling can effectively avoid segmentation ambiguity while not altering the size of the candidate pool or the probability distribution of any token within it.

The main contributions of this paper can be summarized as follows:
\begin{itemize}
    \item \textbf{Analysis of the problems of segmentation ambiguity.} We review the segmentation ambiguity issue faced by existing provably secure steganography methods and analyze why current disambiguating solutions are not applicable to provably secure steganography. 
    \item \textbf{The first provably secure disambiguating linguistic steganography.} We propose a novel solution named SyncPool based on ambiguity pool grouping and synchronous sampling to address information loss and token synchronization issues during steganography, eliminating segmentation ambiguity without altering the distribution.
    \item \textbf{Theoretical proof of steganographic security.}
    We theoretically prove that SyncPool does not alter the probability distribution of the model during steganographic embedding, and from a computational security perspective, we reduce the security of the algorithm to the security of the CSPRNG.

    \item \textbf{Considerable performance of effectiveness and efficiency.} We designed and implemented necessary experiments to demonstrate that our method can completely eliminate decoding errors, and provided analysis on the impact on embedding and time efficiency.

\end{itemize}

The source code of our implementations of SyncPool can be found at \url{https://github.com/7-yaya/SyncPool}.
\section{Background and Related Work}

\subsection{Steganography System and Steganographic Security}

Steganography is a technique that embeds secret messages into objects that closely resemble real, mundane communications, making it impossible for censors to suppress such communications. Steganography is usually illustrated by Simmons' \textit{Prisoners' Problem}~\cite{simmons1984prisoners}. Two prisoners, Alice and Bob, are attempting to communicate with each other over a monitored channel while trying to avoid arousing suspicion from the warden Eve. Therefore, they must find some way to embed the secret message into an ``innocent-looking'' \textit{cover} to obtain a \textit{stego}.

To this end, given a channel distribution $\mathcal{D}$ (alias of $P_c$), the following steganography system (stegosystem) is designed:
\begin{itemize}
    \item Alice uses a probabilistic algorithm $\textsc{Encode}_{\mathcal{D}}$ to embed a secret message $\mathbf{m} \in \{0,1\}^*$ with a shared key $K$ and a channel history $\mathcal{H}$ to obtain a \textit{stego} $\mathbf{s}$ and sends it to Bob.
        \begin{equation}
            \textsc{Encode}_{\mathcal{D}}\left(K, \mathcal{H}, \mathbf{m}\right)=\mathbf{s}\label{emb}. 
        \end{equation}
    \item Bob receives the stego $\mathbf{s}$ through the monitored channel and extracts the message $\mathbf{m}$ through a probabilistic algorithm $\textsc{Decode}_{\mathcal{D}}$ with a shared key $K$ and a channel history $\mathcal{H}$.
        \begin{equation}
            \textsc{Decode}_{\mathcal{D}}\left(K, \mathcal{H}, \mathbf{s}\right)=\mathbf{m}\label{ext}. 
        \end{equation}
    \item Eve needs to judge whether the object transmitted in the channel is innocent or not.
\end{itemize}

There are two common definitions of steganographic security.
Cachin~\cite{cachin1998information} first proposed an information-theoretic model for steganography with passive adversaries. The adversary’s task of distinguishing between an innocent covertext $c$ and a stegotext $s$ containing a secret message is interpreted as a ``hypothesis testing'' problem. The security of a stegosystem can be quantified by the relative entropy (a.k.a. Kullback-Leibler divergence) between the cover distribution $P_c$ and the stego distribution $P_s$,
\begin{equation}
    D_{KL}(P_c||P_s)=\sum_{x\in\mathcal{C}}P_c(x)\log{\frac{P_c(x)}{P_s(x)}}, 
\end{equation}
where $x$ is the object transmitted in the channel with the alphabet $\mathcal{C}$. If $D_{KL}(P_c||P_s) = 0$, the stegosystem is called \textit{perfectly secure}.
Another definition is based on computational complexity theory, proposed independently by Hopper \textit{et al.}~\cite{hopper2002provably} and Katzenbeisser and Petitcolas~\cite{katzenbeisser2002defining}. The stegosystem is called secure against \textit{chosen hiddentext attacks} if all probabilistic polynomial time (PPT) adversaries $\mathcal{A}$’s advantage against the stegosystem
\begin{equation}
    \left|\text{Pr}\left[\mathcal{A}_\mathcal{D}^{\textsc{Encode}_{\mathcal{D}}\left(K,\cdot,\cdot\right)}=1\right]-\text{Pr}\left[\mathcal{A}_\mathcal{D}^{\mathcal{O}_{\mathcal{D}}\left(\cdot,\cdot\right)}=1\right]\right|<\text{negl}\left(\kappa\right), 
\end{equation}
where $\mathcal{O}_{\mathcal{D}}\left(\cdot,\cdot\right)$ is an oracle that can randomly sample from the channel distribution $\mathcal{D}$, $\kappa$ is the security parameter of the shared key $K$ (usually the length of $K$), and $\text{negl}\left(\kappa\right)$ is a negligible function concerning $\kappa$.

\subsection{Provably Secure Neural Linguistic Steganography}

While the concept of provably secure steganography has been around for some time, classical constructions for provably secure steganography either require a random sampling oracle or explicit data distributions~\cite{chen2021distribution,dingDiscopProvablySecure2023a}. 
These conditions are difficult to meet in traditional data environments. It was not until recent years that the rapid advancement of deep generative models provided the possibility of reproducible sampling for provably secure steganography, and the widespread dissemination of generated data provided a favorable environment for concealing steganographic behavior~\cite{chen2018provably}. Researchers have proposed several provably secure linguistic steganography methods~\cite{zhang2021provably,kaptchuk2021meteor,de2022perfectly,dingDiscopProvablySecure2023a,zhang2024provably} based on deep language models. These methods are dedicated to designing message embedding algorithms that are indistinguishable from the normal generation process, i.e., random sampling. 
Here, we briefly introduce several of these methods.
\begin{itemize}
    \item Kaptchuk \textit{et al.}~\cite{kaptchuk2021meteor} introduced the Meteor method, a provably secure steganography approach based on interval reversibility using a random sampling process akin to arithmetic encoding. During the information embedding process, encrypted information transformed into random numbers falls within a probability interval associated with a token, thereby determining the current generated word as that token.
    \item De Witt \textit{et al.}~\cite{de2022perfectly} proposed a provably secure steganography method based on minimum entropy coupling. They view steganography as a coupling established between message distribution and channel distribution and control the sampling of the channel distribution through coupling influenced by a uniform distribution of messages, thereby mapping messages into samples that follow the carrier distribution.
    \item Ding \textit{et al.}~\cite{dingDiscopProvablySecure2023a} utilized the concept of ``sampled distribution'' to express information and presented the Discop method based on distribution copies. This method defines a probability distribution, from which multiple distribution copies are created, and then uses the index values of distribution copies to express messages.
\end{itemize}

The above methods can achieve provably secure linguistic steganography based on autoregressive language models~\cite{brown2020language}. None of them alter the probability distribution of words to be generated by the model during the process of embedding secret messages. 
Apart from the algorithms, the protocol of the covert communication protocol that both parties need to share also includes the same language model, pseudorandom number generator (PRNG), keys, and contexts used for each generation. Specifically:
\begin{itemize}
    \item \textit{Language models,} including pretrained models and tokenizers. Pretrained language models can be obtained from the open-source platform Hugging Face~\cite{lhoest2021datasets}, and models that are downloaded and used more frequently are more suitable as camouflage environments for steganography. Commonly used open-source models include GPT-2~\cite{radford2019language}, LLaMA~\cite{touvron2023llama1}, GLM~\cite{du2021glm}, etc.
    \item A \textit{pseudorandom number generator} (PRNG) is an algorithm for generating a sequence of numbers whose properties approximate the properties of sequences of random numbers. The PRNG-generated sequence is completely determined by the PRNG seed. Provably secure steganography usually requires a cryptographically secure PRNG (CSPRNG). A requirement for a CSPRNG is that an adversary not knowing the seed has only a negligible advantage in distinguishing the generator's output sequence from a random sequence.
    \item \textit{Keys.} In general, steganography involves two keys: an encryption key to encrypt the message before embedding and to decrypt the message after extraction; and a steganograpy key as the CSPRNG seed to initialize the CSPRNG.
    In some cases, the two keys can be one.
    \item \textit{Context.} The same initial context of each generated text needs to be shared, which can be empty for unconditional generation or consist of a specified number of sentences with or without a specific template.
\end{itemize}

\subsection{Segmentation Ambiguity}

Like other traditional generative linguistic steganography methods, provably secure linguistic steganography would also suffers from segmentation ambiguity~\cite{nozaki2022addressing}. 
Assume that the sender Alice and the receiver Bob use one of the above provably secure steganography methods to transmit a stegotext. They share a language model, an initial context, and a key in advance. Alice uses the steganographic encoding algorithm to generate a continuation of the context with the message embedded to obtain a stego. 

The commonly used high-performance Transformer-based language models, represented by the GPT series~\cite{radford2019language,brown2020language,achiam2023gpt} and the LLaMA series~\cite{touvron2023llama1,touvron2023llama2},  mostly utilize \textit{subword tokenization}~\cite{sennrich2015neural,kudo2018subword,kudo2018sentencepiece} to model text. Subword tokenization is a technique in which a word is split into subwords, and these subwords are known as tokens. This technique is used because a generative language model needs to maintain a large vocabulary and complex word structures. The concept behind this is that frequently occurring words should be in the vocabulary whereas rare words are split into frequent subwords. Each token is assigned an ID as a numerical representation of the subword~\footnote{Since the numeric IDs corresponding to tokens differ across different models, in this paper, we directly use the subword mapped by the token to represent the token. In this paper, we use the underscore symbol ``\_'' to indicate a word boundary, which is generally equivalent to a space in English.}. 
For example, the word ``\_unwanted'' might be split into ``\_un'', ``want'', and ``ed''. 
The stego generated by the steganographic encoding algorithm $\textsc{Encode}_{\mathcal{D}}$ is essentially a sequence composed of stego tokens.
The sender must detokenize it using a tokenizer into a stegotext before transmission. The reason is obvious because even in today's prevalent use of generative models and generated text, transmitting tokens directly over a public channel is a highly suspicious behavior. 
For example, if the sender generates two stego tokens mapping to subwords ``\_any'' and ``thing'', the sender needs to detokenize them into the text ``anything'' before sending it to Bob. 
However, the issue is that common words like ``anything'' often exist as independent tokens ``\_anything'' in the model's vocabulary as well.
As a result, a single piece of text can correspond to two or even more different token representations. This phenomenon is referred to as \textit{segmentation ambiguity}.
To replicate Alice's generation process and extract the message using $\textsc{Decode}_{\mathcal{D}}$, Bob must retokenize the received text into tokens. 
But since both ``\_any'' and ``\_anything'' exist in the candidate pool, Bob cannot determine which token the sender embedded the message into.
Clearly, the differing tokenizations could result in the incorrect extraction of the message. 

Currently, all provably secure steganography methods are ambiguity-unaware. The receiver cannot determine whether they have successfully obtained the token sequence generated by the sender or if they have misunderstood the secret message the sender intended to convey, which severely impacts the usability of provably secure steganography.

\subsection{Disambiguation Algorithm}

In the past two years, solutions have emerged to address segmentation ambiguity in traditional generative linguistic steganography. 

\subsubsection{Basic Solution}

Nozaki and Murawaki~\cite{nozaki2022addressing} proposed a simple disambiguating approach, which removes the tokens whose mapping subwords are prefixes of others during every generation and extraction step. 
This process ensures that any token sent by the sender is uniquely extractable for the receiver. 

\subsubsection{MWIS-based Solution}

Yan \textit{et al.}~\cite{yan2023secure} considered the influence of removing candidate words on the probability distributions and decided to process only if candidate-level ambiguity occured. Their solution identifies the maximum-weight independent set (MWIS) in the candidate pool to reduce probability distortion.

Regardless of the method, modifications to the candidate pool evidently alter the probability distribution, resulting in significant discrepancies from the distribution of normal generation processes. Therefore, these existing token-removal-based solutions for addressing segmentation ambiguity are not applicable to provably secure steganography.



\begin{figure*}[!t]
    \centering
    \includegraphics[width=\textwidth]{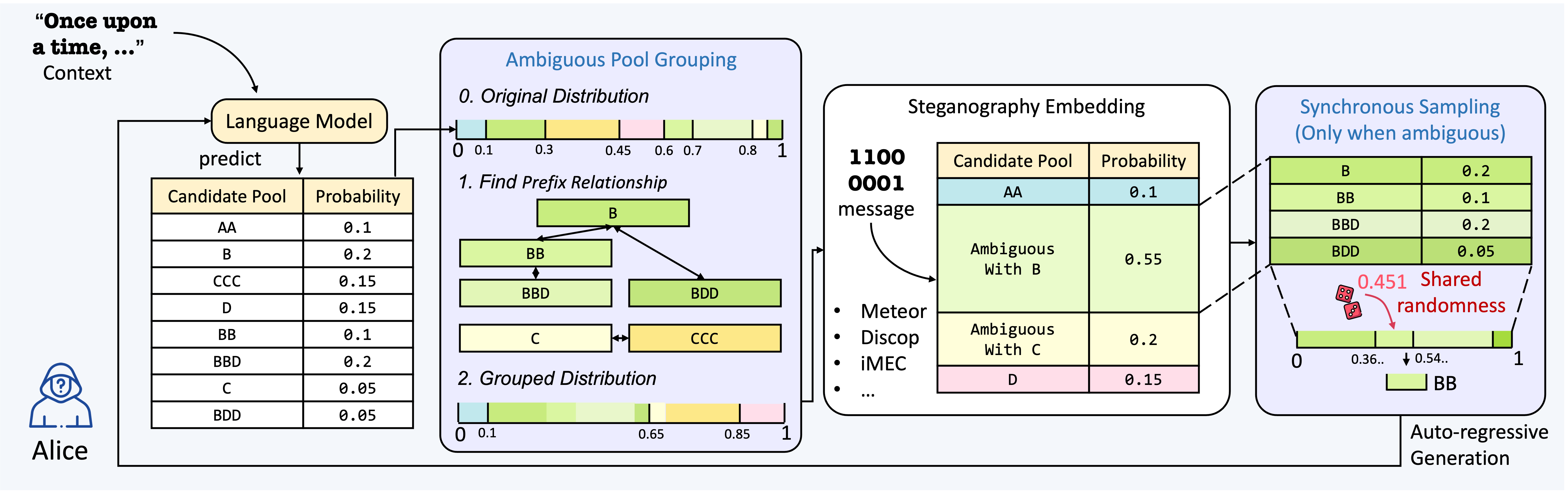}
    \caption{The provably secure disambiguating linguistic steganography consists of the existing provably secure steganography method and our proposed distribution-preserving disambiguation method SyncPool. We group the original probability distribution by prefix relationships, use the existing provably secure steganography method, e.g. Discop, to embed the message on the grouped distribution, and perform synchronized random sampling from the ambiguity pools using a shared random number between the sender and receiver to ensure unique message extraction. The probability distribution corresponding to the sampled tokens remains unchanged before and after implementing ambiguity elimination.}
    \label{fig_flow}
\end{figure*}

\section{Proposed Method}\label{sec_3}

As analyzed above, previous attempts at disambiguating introduced damage to the original probability distribution while removing potentially ambiguous tokens. An adversary can easily obtain a nonnegligible advantage in distinguishing the cover distributions and the stego distributions. 
The ideal method must satisfy the following criteria: 
(1) 
the disambiguation algorithm should not alter the model's original probability distribution, all tokens in the candidate pool predicted by the language model should be retained, and each token should maintain its original probability; 
(2) the disambiguation algorithm should be easily transferable to any generative linguistic steganographic method based on autoregressive models without being specialized for a specific embedding and extraction algorithm. 
To address this problem, in this section, we present a disambiguation method named \textbf{SyncPool}, which can help both parties in communication eliminate segmentation ambiguity in steganography without altering the distribution.

Our disambiguation method SyncPool primarily consists of two modules: \textit{ambiguity pool grouping} and \textit{synchronous sampling}. 
The first module acts before steganographic embedding and groups tokens according to prefix relationships, which eliminates the uncertainty between ambiguous tokens and avoids information loss during detokenization.
The second module works after the steganography algorithm selects an ambiguity pool to help both the sender and the receiver achieve synchronous sampling from the pool.
The disambiguating steganography embedding process is shown in \autoref{fig_flow}.
In this method, the number of sampleable tokens and the probability of each token remain unchanged compared to the original probability distribution. 

Next, we will first define the original candidate pool and distribution for steganography and then provide a detailed description of each part of the disambiguation algorithm.

\subsection{Candidate Pool with Original Distribution}

During normal generation, the language model can predict the probability distribution of the next token $x_t$ given the previous context $x_{<t}$ in the whole vocabulary $\Sigma$:
\begin{equation}
    \text{Pr}\left[x_t|x_{<t}\right] = \left[p_1, p_2, \dots, p_{\left|\Sigma\right|}\right], 
\end{equation}
where $p_1$ through $p_{\left|\Sigma\right|}$, respectively represent the probabilities of tokens $w_1$ through $w_{\left|\Sigma\right|}$ in the vocabulary, holding that $\sum_{i=1}^{\left|\Sigma\right|}p_i=1$.

Then, the model needs to sample a token from the distribution mentioned above and add it to the generated sequence. There are primary two sampling methods: \textit{greedy sampling} and \textit{random sampling}. In most cases, users of the generative language model desire the model's output to exhibit variability and creativity while maintaining a level of controlled randomness. Therefore, to ensure that the model's output meets specific requirements or matches the desired styles, three crucial parameters are introduced during the random sampling process. They are temperature, top-$p$, and top-$k$. 

The temperature influences the shape of the probability distribution that the model calculates for the next token rather than limiting the token selection. 
Top-$p$, i.e., nucleus sampling~\cite{holtzman2019curious}, allows for dynamic control of the number of tokens considered, leading to different levels of diversity in the generated text, while top-$k$ provides controlled randomness by considering a fixed number of top probable tokens. 
Three of them could change the distributions or candidate pools of $x_t$. 
It is common to use these parameters to control random sampling in the normal sampling process of innocent users.
Therefore, in provably secure steganography, there should be no restrictions on the use of these parameters. Additionally, reducing the size of the candidate pool using the top-$p$ and top-$k$ methods significantly reduces the computational overhead of both random sampling and generative steganography. 

Therefore, in this paper, we define the \textbf{original distribution} as the probability after applying these probability processing methods and the \textbf{original candidate pool} as the remaining available tokens after the top-$k$ or top-$p$ stage, which does not contradict the behavioral security pursued by steganography.

\begin{definition}
    The original distribution of the language model before steganography refers to the probability distribution composed of the remaining tokens after probability processing. Let $V$ denote the remaining \textbf{candidate pool}, $V=\left[w_1, w_2, \dots, w_{\left|V\right|}\right]$, $\left|V\right|\leq\left|\Sigma\right|$. The \textbf{original distribution} is $P_c=\left[p_1, p_2, \dots, p_{\left|V\right|}\right]$, where $\sum_{i=1}^{\left|V\right|}p_i\leq1$. 
\end{definition}

Provable secure steganography requires that steganography does not alter the original distribution predicted by the model. Therefore, the candidate pool and distribution after disambiguation must also remain consistent with $V$ and $P_c$. 

\subsection{Ambiguity Pool}

Like the existing disambiguation methods, our method also needs to identify all potential candidate words that could cause ambiguity. 
The \textbf{prefix relationship} between candidate words is a necessary but not sufficient condition for segmentation ambiguity to occur. 

\begin{definition}
    In the candidate pool, tokens exhibit \textbf{prefix relationships}, meaning that for a token $w_i$ in the candidate pool, its mapping subword either serves as a prefix for another token $w_j$ or another token's mapping subword is a prefix of this token's.  
\end{definition}

Formally, a prefix relationship can be denoted as:
\begin{equation}
    \textsc{prefix}\left(w_i, w_j\right) = 
    \begin{cases}
        1, & w_i \text{ is a prefix of } w_j , \\
        1, & w_j \text{ is a prefix of } w_i , \\
        0, & \text{otherwise} .
    \end{cases}  
\end{equation}
where tokens $w_i$ and $w_j$ are in the candidate pool $V^{(t)}$. $V^{(t)}$ represents the remaining candidate pool during the $t$-th time step of the generative model's sampling process. All occurrences of $(t)$ in the following text have the same meaning, representing the time step applicable to the corresponding data. In the following algorithm description, we use $\textsc{prefix}\left(w_i, w_j\right)=1$ to represent the presence of a prefix relationship between two tokens.

\begin{figure}[htbp]
    \centering
    \subfloat[$D$ /wo prefix relationships]{\includegraphics[width=0.4\columnwidth]{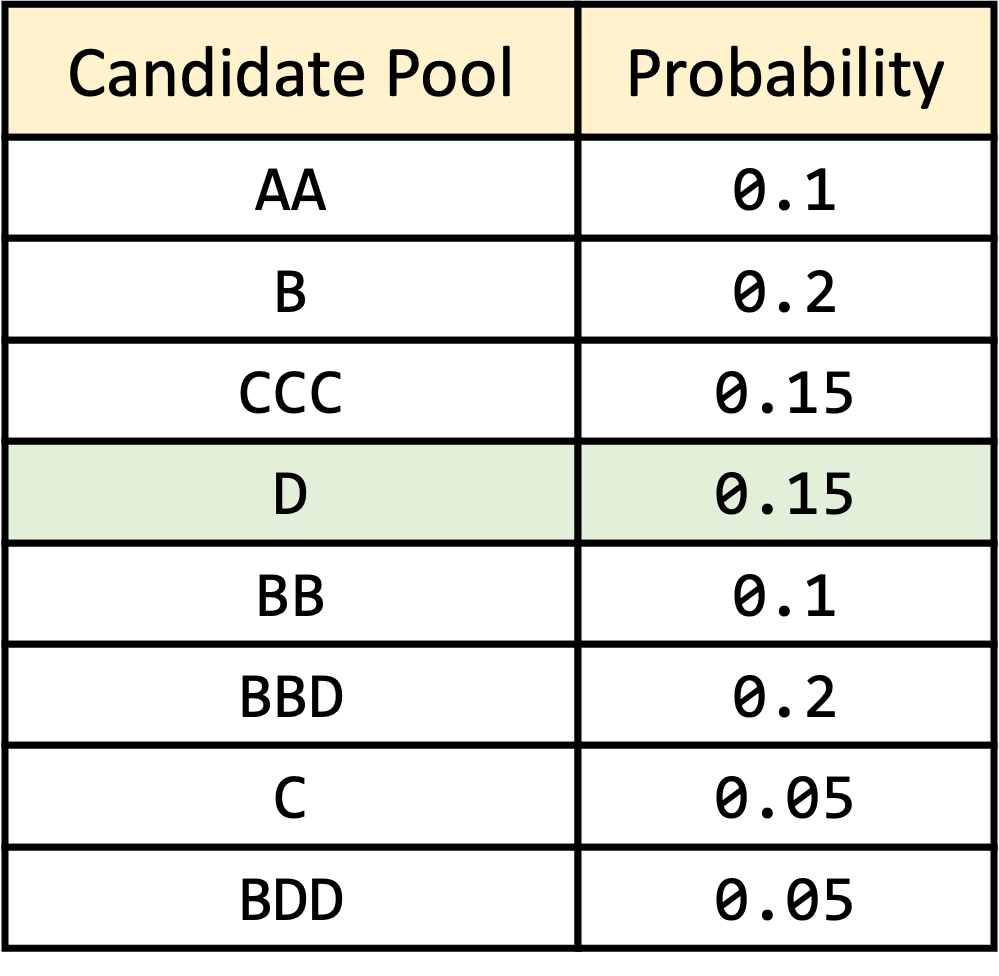}}%
    \label{fig_first_case_1}
    \hfil
    \subfloat[$BB$ /w prefix relationships]{\includegraphics[width=0.4\columnwidth]{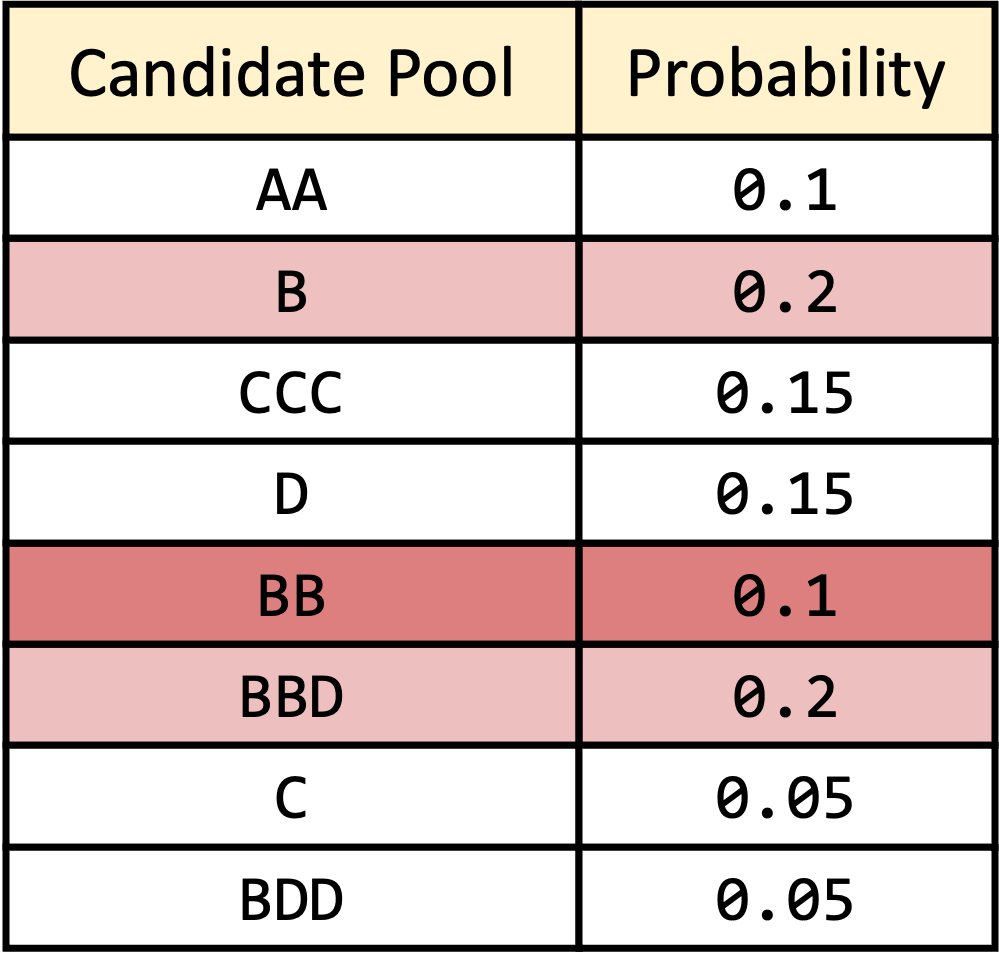}}%
    \label{fig_first_case_2}
    \caption{Different cases about prefix relationship.}
    \label{fig_prefix}
\end{figure}

\autoref{fig_prefix} illustrates two scenarios: one where a selected token has no prefix relationship with other tokens in the candidate pool and another where it does. Given a candidate pool of the top-$8$ for an example, $V=[AA,B,CCC,D,BB,BBD,C,BDD]$. When the selected token is $D$, there is no prefix relationship; when it is $BB$, two other tokens, it holds that $\textsc{prefix}\left(B, BB\right)=1$ and $\textsc{prefix}\left(B, BBD\right)=1$.

The presence of prefix relationships causes two tokens corresponding to different messages in steganography to convert into a entirely identical text for the reciever after detokenization.
The disappearance of uncertainty implies a loss of information, meaning the embedded message in steganography cannot be extracted. 
This is the reason why segmentation ambiguity leads to decoding errors.

The previous disambiguation algorithms removed certain tokens to ensure that there were no prefix relationships in the candidate pool, which inevitably altered the probability distribution of tokens.
To avoid compromising steganographic security, we first address the information loss caused by ambiguous tokens.
The method we implemented in this paper is to establish new groups for tokens that have prefix relationships. We name this kind of group consisting of ambiguous tokens as an \textbf{ambiguity pool}. 
In this paper, establishing ambiguity pools that can synchronize the sender and the receiver for a given candidate pool is the key to our disambiguation algorithm. 

\begin{algorithm}[t]
\caption{$\textsc{Ambiguity}\left(\Vt, P_\wt\right)$: Construct an Ambiguity Pool on Candidate Pool $\Vt$ and its distribution $P_\wt$.}\label{algo_ambiguouspool}

\SetAlgoLined
\small
\KwIn{Candidate pool $\Vt=\left[w_1, w_2, \dots, w_{\left|V\right|}\right]$, candidate distribution $P_\wt=\left[p_1, p_2, \dots, p_{\left|V\right|}\right]$;} 
\KwOut{Ambiguity pool $V_{amb}^{(t)}$, ambiguity distribution $P_{amb}^{(t)}$;}

$\Vt,P_\wt\leftarrow\textsc{Sort}\left(\Vt,P_\wt\right)$\;
$V_{amb}^{(t)}\leftarrow\emptyset$\text{,\ }$P_{amb}^{(t)}\leftarrow P_\wt$\;

$j\leftarrow1,\ $
$v_{amb,j}\leftarrow[w_j],$
$p_{amb,j}\leftarrow [p_j]$\;

\For{$w_i$ in $\left[w_2, \dots, w_{\left|V\right|}\right]$}{
    \eIf{$\textsc{prefix}\left(w_j, w_i\right)$}{
        $v_{amb,j}.append(w_i)$\;
        $p_{amb,j}.append(p_i)$\;
    }{
        $V_{amb}^{(t)}.append(v_{amb,j})$\;
        $P_{amb}^{(t)}.append(p_{amb,j})$\;
        $j\leftarrow j+1$\;
        $w_{j}\leftarrow w_i$\;
        $v_{amb,j}\leftarrow[w_j]$\;
        $p_{amb,j}\leftarrow p_j$\;
    }
}

\textbf{return}
\end{algorithm}

The specific algorithm for constructing the ambiguity pools on a given candidate pool is shown in \autoref{algo_ambiguouspool}. 
We first sort all the tokens in the pool according to the character order of their corresponding subwords. After sorting, we check whether each token shares a common prefix as a token with the previous token. If so, we merge it into a group. If not, we start merging the next group. 
The tokens in each merged group jointly form an ambiguity pool $v_{amb}$. 
A new distribution is established, denoted as $p_{amb}$, for these elements in $v_{amb}$ such that the probability of each token remains consistent with the original probability distribution. 
All ambiguity pools collectively form the candidate pool after grouping. We refer to it as $V_{amb}^{(t)}$, where t is the current generation time step, with its corresponding probability distribution denoted as $P_{amb}^{(t)}$.
The process of merging ambiguous tokens and constructing ambiguity pools can be denoted as: 
\begin{equation}
    V_{amb}^{(t)}, P_{amb}^{(t)}\leftarrow\textsc{Ambiguity}\left(\Vt, P_w^{(t)}\right).  
\end{equation}
For an ambiguity pool $v_{amb,j}$ in $V_{amb}^{(t)}$, we define its representative token as the first token in the ambiguity pool, i.e., the token whose mapping subword is the shortest, denoted as $w_{amb,j}$. We refer to $v_{amb,j}$ as an ambiguity pool with $w_{amb,j}$, denoted as $v_{amb,j}^{w_{amb,j}}$.


Using the distribution in~\autoref{fig_prefix} as an example, the original candidate pool can be sorted to 
\begin{equation}
    V' = [AA,B,BB,BBD,BDD,C,CCC,D].  
\end{equation}
Then, tokens that have prefix relationships are merged, and a new distribution is established:
\begin{equation}
    V_{amb}=[AA,[B,BB,BBD,BDD],[C,CCC],D],  
\end{equation}
where $\left|V_{amb}\right|=4$. The groups $[B,BB,BBD,BDD]$ and $[C,CCC]$ are called the \textit{ambiguity pools with $B$} and \textit{with $C$}, respectively.
Notably, under the ambiguity pool partitioning scheme in this context, not all tokens within an ambiguity pool exhibit a prefix relationship. For example, in the ambiguity pool with $B$ mentioned above, there is no prefix relationship between $BB$ and $BDD$ as defined previously; in other words, $BB$ is not a prefix of $BDD$, and $BDD$ is not a prefix of $BB$. The same logic applies to $BBD$ and $BDD$. However, they all share a prefix relationship with $B$ and are thus grouped together.

After grouping is completed, we transform the objects for embedding and extraction in the steganographic algorithm into an ambiguity pool. This ensures that regardless of which ambiguous token is selected from the ambiguity pool, the same information will be expressed, thus avoiding information loss during the detokenization.

\subsection{Synchronous Random Sampling}

While the partitioning of ambiguity pools resolves the issue of information loss between detokenization and tokenization, it does not aid in synchronizing tokens between the sender and the receiver. 
Due to the autoregressive nature of the model, if the sender and receiver have different understandings of segmentation for the same text, it may not affect the extraction of messages in the current step but will affect the probabilities predicted by the model for all subsequent generation steps.
Therefore, we design a new mechanism that enables the sender and receiver to \textit{synchronize the selection of tokens from ambiguity pools} using a preshared random number.
We name this stage \textbf{synchronous random sampling}. 

Formally, let $v_{amb}^{(t)}$ be the ambiguity pools with $w_{amb}^{(t)}$ for step $t$, and let $p_{amb}^{(t)}$ be its corresponding probability distribution. 
We generate a random number that is shared between the sender and receiver via a synchronous CSPRNG, denoted as $\text{CSPRNG}_{sync}$. As long as the sender and receiver share an initial seed, or a symmetric key, they can obtain a series of synchronized pseudorandom numbers $\mathbf{r}=\left\{r^{(0)}, r^{(1)}, \dots\right\}$, which follows a uniform distribution on $\left[0,1\right)$, i.e., $r^{(t)}\sim U\left[0,1\right)$. 
At each time step $t$, each token in the ambiguity pool $v_{amb}^{(t)}$ is allocated to the left-closed, right-open interval $\left[0,1\right)$ according to $p_{amb}^{(t)}$. Then, we consume a pseudorandom number $r^{(t)}$ and select the token corresponding to the interval into which $r^{(t)}$ falls, denoted as the synchronous token $w_{sync}$:
\begin{equation}
    w_{sync}^{(t)} \leftarrow \textsc{SyncSample}\left(p_{amb}^{(t)}, r^{(t)}\right).  
\end{equation}
As long as both the sender and receiver adhere to the same ambiguity pool construction method, the receiver can determine which specific token the sender has chosen from the ambiguity pool in each generation step.
Because the same shared random number corresponds to only one token, ambiguity in segmentation naturally disappears.


\begin{algorithm}[t]
\SetAlgoLined
\caption{$\textsc{Embedding}$}\label{algo_embedding}
\small
\KwIn{Context $C$, $\text{CSPRNG}_{steg}$, $\text{CSPRNG}_{sync}$, Language Model $\mathcal{M}$, Tokenizer $\mathcal{T}$, Seed $K$, Message $\mathbf{m}$}
\KwOut{Stegotext $S$}

\begin{minipage}[t]{0.3\columnwidth}
$x\leftarrow\emptyset$\;
\end{minipage}
\hfill
\begin{minipage}[t]{0.5\columnwidth}
\raggedleft
\texttt{// Stego Tokens}
\end{minipage}

$\text{CSPRNG}_{steg}.\textsc{SetSeed}\left(K\right)$\;
$\text{CSPRNG}_{sync}.\textsc{SetSeed}\left(K\right)$\;
$t\leftarrow1$

\While{\textbf{not} the end of $\mathbf{m}$}{
    \begin{minipage}[t]{0.5\columnwidth}
    $\Vt, P_w^{(t)}\leftarrow \mathcal{M}\left(C\right)$\;
    \end{minipage}
    \hfill
    \begin{minipage}[t]{0.3\columnwidth}
    \raggedleft
    \texttt{// predict}
    \end{minipage}
    $V_{amb}^{(t)}, P_{amb}^{(t)}\leftarrow\textsc{Ambiguity}\left(\Vt, P_w^{(t)}\right)$\;

    $v_{amb}^{(t)}\leftarrow\textsc{Encode}_{P_{amb}^{(t)}}\left(\text{CSPRNG}_{steg}(K)^{(t)}, \mathbf{m}\right)$\;
    
    $\mathbf{m} \leftarrow \mathbf{m}\left[\textsc{EmbedNum}\left(v_{amb}^{(t)},P_{amb}^{(t)}\right):\right]$\;

    \eIf{\textbf{not} ambiguous}{
	$x_t \leftarrow \wt_{amb}$\;
    }{
	$p_{amb}^{(t)}\leftarrow\frac{p_{amb}^{(t)}}{\sum_{k}^{\left|v_{amb}^{(t)}\right|}p_{amb}^{k}}$\;
 
        $w_{sync}^{(t)} \leftarrow \textsc{SyncSample}\left(p_{amb}^{(t)}, \text{CSPRNG}_{sync}(K)^{(t)}\right)$\;
        $x_t \leftarrow w_{sync}^{(t)}$\;

    }
    $x\textsc{.append}(x_t)$\;
    $C\leftarrow C\parallel x_t$\;
    $t\leftarrow t+1$\;
}
$S\leftarrow \mathcal{T}.\textsc{Decode}\left(x\right)$\;
\textbf{return}
\end{algorithm}

\begin{algorithm}[t]
\SetAlgoLined
\caption{$\textsc{Extraction}$}\label{algo_extraction}
\small
\KwIn{Stegotext $S$, Context $C$, $\text{CSPRNG}_{steg}$, $\text{CSPRNG}_{sync}$, Language Model $\mathcal{M}$, Tokenizer $\mathcal{T}$, Seed $K$}
\KwOut{Message $\mathbf{m}$}
$\mathbf{m}\leftarrow\text{``\ ''}$\;
$x\leftarrow\emptyset$\;
$\text{CSPRNG}_{steg}.\textsc{SetSeed}\left(K\right)$\;
$\text{CSPRNG}_{sync}.\textsc{SetSeed}\left(K\right)$\;
$t\leftarrow1$\;

\While{\textbf{not} the end of $S$}{
    \begin{minipage}[t]{0.5\columnwidth}
    $\Vt, P_w^{(t)}\leftarrow \mathcal{M}\left(C\right)$\;
    \end{minipage}
    \hfill
    \begin{minipage}[t]{0.3\columnwidth}
    \raggedleft
    \texttt{// predict}
    \end{minipage}

    $\wt\leftarrow\textsc{PrefixToken}\left(S,\Vt\right)$\;

    $V_{amb}^{(t)},P_{amb}^{(t)}\leftarrow\textsc{Ambiguity}\left(\Vt, P_w^{(t)}\right)$\;

    \For{$v_{amb,j}$ in $V_{amb}^{(t)}$}{
        \If{$\wt$ in $v_{amb,j}$}{
            $v_{amb}^{(t)}\leftarrow v_{amb,j}$\;
        }
    }
    $\mathbf{m}_t\leftarrow\textsc{Decode}_{P_{amb}^{(t)}}\left(\text{CSPRNG}_{steg}(K)^{(t)}, v_{amb}^{(t)}\right)$\;
    
    \eIf{\textbf{not} ambiguous}{
	$x_t \leftarrow \wt$\;
 
    }{
        $w_{sync}^{(t)} \leftarrow \textsc{SyncSample}\left(p_{amb}^{(t)}, \text{CSPRNG}_{sync}(K)^{(t)}\right)$\;
        $x_t \leftarrow w_{sync}^{(t)}$\;
    }

    $x\textsc{.append}(x_t)$\;
    $C\leftarrow C\parallel x_t$\;
    
    $S\leftarrow S\left[\textsc{Len}\left(\mathcal{T}.\textsc{Decode}\left(x_t\right)\right):\right]$\;
    $\mathbf{m}\leftarrow \mathbf{m}\parallel \mathbf{m}_t$\;

    $t\leftarrow t+1$\;
    
}
\textbf{return}
\end{algorithm}

Next, we will describe the complete process of provably secure disambiguating steganography, including steganographic embedding and steganographic extraction.

\subsection{Embedding}

The main loop of the provably secure steganographic message embedding process with our proposed disambiguation algorithm is shown in \autoref{algo_embedding}.
At the $t$-th step of generation, the sender first needs to group and construct ambiguity pools for tokens that may cause ambiguity based on the current candidate pool and distribution. 
Then the steganography embedding algorithm provides an ambiguity pool to be output based on the distribution of ambiguity pools $P_{amb}^{(t)}$, a random number provided by the $\text{CSPRNG}_{steg}$ given the key $K$, and the message bit sequence $\mathbf{m}$, 
\begin{equation}
v_{amb}^{(t)}\leftarrow\textsc{Encode}_{P_\wt}\left(\text{CSPRNG}_{steg}(K)^{(t)}, \mathbf{m}\right).  
\end{equation}
If there is only one token in the pool $v_{amb}^{(t)}$, it can be added to the stego sequence, i.e., $x_t\leftarrow w_{amb}^{(t)}$, and the steganographic message embedding and generation process finishes normally. 

However, in the event of ambiguity, i.e., $v_{amb}^{(t)}$ contains more than one element, the steganographic embedding process does not select a specific token, but rather selects a group of tokens that may cause ambiguity. 
The information that can be embedded is determined by the probability of this group. 
At this point, we employ the synchronous sampling method proposed in the previous section for token synchronization. 
The synchronized ambiguous token $w_{sync}^{(t)}$ is added to the stego token sequence.

Finally, the sender needs to detokenize the stego tokens into stegotext before transmission.

\subsection{Extraction}

The main loop of the provably secure steganographic message extraction process with our proposed disambiguation algorithm is shown in \autoref{algo_extraction}. 
During extraction, unlike the previous ambiguity-unaware provably secure steganography, the receiver needs to first abandon using the off-the-shelf tokenizer to tokenize the stegotext directly. 
At each time step of the autoregressive generation process, the receiver selects a token $\wt$ from the candidate pool $\Vt$, while $\wt$ is a prefix of the remaining part of the detokenized stegotext $S$. 
The receiver then locates the ambiguity pool $v_{amb}^{(t)}$ corresponding to the selected token $\wt$ and extracts the message based on the distribution of the ambiguity pool. Next, they employ the same synchronization sampling as the sender to ensure accurate segmentation.
This process is repeated until the stegotext is fully extracted.

\subsection{Complexity}

Let $V$ denote the candidate pool and $|V|$ denote the size of the given pool. 
When SyncPool is used, the disambiguation algorithm performs an extra sorting operation and a traversal of all tokens in the candidate pool to compare common prefixes. The complexity of the sorting operation is $O(|V|\log(|V|))$, and the complexity for comparing common prefixes is $O(|V|)$. 
After this, the size of the candidate pool is reduced to $|V_{amb}|$. 
The complexity of steganographic embedding of the candidate pool changes accordingly, as determined by the steganographic algorithm used.

\subsection{Proof of Security}

From the perspective of computational security, we prove that SyncPool is a secure method for disambiguating steganography.
By definition, steganography is secure 
aganist chosen hiddentext attacks, if all probabilistic polynomial time (PPT) adversaries $\mathcal{A}$’s advantage against the stegosystem 
\begin{equation}
    \left|\text{Pr}\left[\mathcal{A}_{\mathcal{D}}({x_s})=1\right]-\text{Pr}\left[\mathcal{A}_{\mathcal{D}}({x_c})=1\right]\right|<\text{negl}\left(\kappa\right).
\end{equation}
where $x_s$ is the disambiguating steganographic stegotext and $x_c$ is the normally generated covertext, $\kappa$ is the security parameter of the shared key $K$ (usually the length of $K$), and $\text{negl}\left(k\right)$ is a negligible function concerning $\kappa$.

We prove the statement using a proof by contradiction. First, assume that the disambiguating steganographic stegotext $x_s$ and the normally generated covertext $x_c$ are distinguishable, meaning
\begin{equation}\label{eq_assume1}
    \left|\text{Pr}\left[\mathcal{A}_{\mathcal{D}}({x_s})=1\right]-\text{Pr}\left[\mathcal{A}_{\mathcal{D}}({x_c})=1\right]\right|=\delta, 
\end{equation}
where $\delta$ is non-negligible with respect to the key $K$. Considering the tokens generated at each step, we have
\begin{equation}\label{eq_assume2}
    \left|\text{Pr}\left[\mathcal{A}_{P_c}({w_s})=1\right]-\text{Pr}\left[\mathcal{A}_{P_c}({w_c})=1\right]\right|=\delta, 
\end{equation}
In our method, a token to be added to the stego sequence undergoes two sampling processes. 
The first sampling occurs within the distribution $P_{amb}$ formed after grouping by the ambiguity pool, controlled by the steganographic embedding algorithm and driven by the secret message. The result is a certain ambiguous pool $v_{amb}$, consisting of one or more ambiguous tokens.
The second sampling occurs within $v_{amb}$, driven by pseudorandom numbers $r$ generated by a CSPRNG.
For simplicity, we use $\mathcal{S}(r, P)$ to represent sampling from $P$ using $r$, and $\mathcal{E}(m, P)$ to represent the steganography algorithm $\textsc{Encode}_{P}(\cdot,m)$, where $m$ is the secret message encrypted using a cryptographic algorithm. Therefore, the adversary's judgement on $w_s$ can be expressed using the law of total probability as 
\begin{align}
    \nonumber
    &\text{Pr}\left[\mathcal{A}_{P_c}({w_s})=1\right]\\\nonumber
    &= \text{Pr}\left[\mathcal{A}_{P_c}({w_s})=1\mid \mathcal{A}_{P_c}({v_{amb}^{w_s}})=1\right] \text{Pr}\left[\mathcal{A}_{P_c}({v_{amb}^{w_s}})=1\right]\\
    &+\text{Pr}\left[\mathcal{A}_{P_c}({w_s})=1\mid \mathcal{A}_{P_c}({v_{amb}^{w_s}})=0\right]  \text{Pr}\left[\mathcal{A}_{P_c}({v_{amb}^{w_s}})=0\right],
\end{align}
where $w_s=\mathcal{S}(r,p_{amb}^{w_s})$, $v_{amb}^{w_s}=\mathcal{E}(m,P_{amb})$.
At the same time, as the probabilities of each token remain unchanged before and after being grouped, a single normal generation of samples is also equivalent to two separate sampling processes using random numbers.
The adversary's judgment on $w_c$ can also be expressed as
\begin{align}
    \nonumber
    &\text{Pr}\left[\mathcal{A}_{P_c}({w_c})=1\right]\\\nonumber
    &= \text{Pr}\left[\mathcal{A}_{P_c}({w_c})=1\mid \mathcal{A}_{P_c}({v_{
amb}^{w_c}})=1\right] \text{Pr}\left[\mathcal{A}_{P_c}({v_{amb}^{w_c}})=1\right]\\
    &+\text{Pr}\left[\mathcal{A}_{P_c}({w_c})=1\mid \mathcal{A}_{P_c}({v_{amb}^{w_c}})=0\right]  \text{Pr}\left[\mathcal{A}_{P_c}({
v_{amb}^{w_c}})=0\right].
\end{align}
We also have $v_{amb}^{w_c}=\mathcal{S}(r_1,P_{amb})$ and $w_c=\mathcal{S}(r_2,p_{amb})$, where $r_1$ and $r_2$ are independent random variables corresponding to the equivalent two-stage sampling, each uniformly distributed, i.e. $r_1,r_2\stackrel{\text{i.i.d.}}{\sim} U\left[0,1\right)$.

Upon comparison, it can be observed that to satisfy Eq.~\eqref{eq_assume2}, it must be the condition that: the adversary can distinguish between $
\mathcal{E}(m,P_{amb})$ and $\mathcal{S}(r_1,P_{amb})$ in polynomial time, or the adversary can distinguish between $\mathcal{S}(r,p_{amb}^{w_s})$ and $\mathcal{S}(r_2,p_{amb})$ in polynomial time.
Note that the provable security of the steganography algorithm we employ guarantees that the PPT adversary cannot have a negligible advantage $\delta^{'}$ in distinguishing between $\mathcal{E}(m,P_{amb})$ and $\mathcal{S}(r_1,P_{amb})$, that is
\begin{equation}
    \left|\text{Pr}\left[\mathcal{A}_{P_c}({v_{amb}^{w_c}})=1\right]-\text{Pr}\left[\mathcal{A}_{P_c}({v_{amb}^{w_c}})=1\right]\right|<\text{negl}\left(\kappa\right).
\end{equation}
Additionally, we use a CSPRNG in our synchronous sampling function, implying that the generated $r$ as a pseudorandom sequence cannot be distinguished from a truly random sequence in polynomial time.
Both conditions cannot be satisfied.
Therefore, Eq.~\eqref{eq_assume2} is not valid, leading us back to the initial assumption that Eq.~\eqref{eq_assume1} is not valid.

Hence, any computationally secure steganographic method with SyncPool is a computationally secure disambiguating steganography. $\qed$

\section{Experiments}
In this section, we conduct experiments to evaluate the performance of our proposed disambiguating algorithm in terms of effectiveness, security and efficiency, and compare our algorithm with baselines attempting to achieve ambiguity-aware provably secure linguistic steganography.

\begin{table*}[!t]
\centering
\caption{Quantitative comparison with the previous disambiguating methods on Discop using LLaMA2 and English context.}\label{tab_exp_en}
\resizebox{0.9\textwidth}{!}{
\begin{tabular}{@{}cc|ccccccccc@{}}
\toprule
\multicolumn{2}{c|}{Method} & $k$ & Ave PPL & Ave KLD $\downarrow$ & Max KLD $\downarrow$ & \begin{tabular}[c]{@{}c@{}}Capacity $\uparrow$\\ (bits/token)\end{tabular} & Utilization $\uparrow$ & \begin{tabular}[c]{@{}c@{}}Total Time $\downarrow$\\ ($\times 10^{-2}$ seconds)\end{tabular} & \begin{tabular}[c]{@{}c@{}}Ave Time $\downarrow$\\ ($\times 10^{-6}$ seconds/bit)\end{tabular} & \begin{tabular}[c]{@{}c@{}}Total Error $\downarrow$\\ (\%)\end{tabular} \\ \midrule
\multicolumn{2}{c|}{\multirow{5}{*}{\begin{tabular}[c]{@{}c@{}}Discop\end{tabular}}} & 16 & 3.27 & 0 & 0 & 1.30 & 0.84 & 1.96 & 1.50 & 1.81 \\
\multicolumn{2}{c|}{} & 32 & 3.50 & 0 & 0 & 1.44 & 0.85 & 2.25 & 1.56 & 2.80 \\
\multicolumn{2}{c|}{} & 64 & 3.88 & 0 & 0 & 1.58 & 0.85 & 3.24 &  2.05 & 2.74 \\
\multicolumn{2}{c|}{} & 128 & 4.20 & 0 & 0 & 1.70 & 0.87 & 4.31 & 2.53 & 2.50 \\
\multicolumn{2}{c|}{} & 256 & 4.55 & 0 & 0 & 1.81 & 0.88 & 7.88 & 4.35 & 2.67 \\ \midrule
\multicolumn{1}{c|}{\multirow{15}{*}{\begin{tabular}[c]{@{}c@{}}Discop +\end{tabular}}} & \multirow{5}{*}{Basic} & 16 & 12.13 & 30.25 & 98.63 & 1.58 & 0.86 & 4.23 & 2.60 & 0 \\
\multicolumn{1}{c|}{} & \multicolumn{1}{c|}{} & 32 & 18.75 & 37.34 & 98.64 & 1.93 & 0.86 & 5.73 & 2.74 & 0 \\
\multicolumn{1}{c|}{} & \multicolumn{1}{c|}{} & 64 & 29.67 & 44.02 & 98.63 & 2.38 & 0.86 & 10.39 & 4.29 & 0 \\
\multicolumn{1}{c|}{} & \multicolumn{1}{c|}{} & 128 & 42.26 & 49.10 & 98.63 & 2.80 & 0.85 & 25.39 & 8.92 & 0 \\
\multicolumn{1}{c|}{} & \multicolumn{1}{c|}{} & 256 & 63.73 & 53.17 & 98.63 & 3.18 & 0.87 & 79.60 & 24.56 & 0 \\ \cmidrule(l){2-11} 
\multicolumn{1}{c|}{} & \multirow{5}{*}{MWIS} & 16 & 2.64 & 2.72 & 64.64 & 1.01 & 0.77 & 13.32 & 11.61 & 0 \\
\multicolumn{1}{c|}{} & \multicolumn{1}{c|}{} & 32 & 2.83 & 2.94 & 71.56 & 1.12 & 0.76 & 20.39 & 15.53 & 0 \\
\multicolumn{1}{c|}{} & \multicolumn{1}{c|}{} & 64 & 3.04 & 3.41 & 72.96 & 1.22 & 0.77 & 37.56 & 26.79 & 0 \\
\multicolumn{1}{c|}{} & \multicolumn{1}{c|}{} & 128 & 3.22 & 3.58 & 68.56 & 1.29 & 0.76 & 133.83 & 85.04 & 0 \\
\multicolumn{1}{c|}{} & \multicolumn{1}{c|}{} & 256 & 3.52 & 3.79 & 72.34 & 1.38 & 0.77 & 1182.07 & 757.59 & 0 \\ \cmidrule(l){2-11} 
\multicolumn{1}{c|}{} & \multirow{5}{*}{SyncPool} & 16 & 3.19 & \textbf{0} & \textbf{0} & 1.00 & 0.66 & 3.90 & 3.85 & 0 \\
\multicolumn{1}{c|}{} & \multicolumn{1}{c|}{} & 32 & 3.74 & \textbf{0} & \textbf{0} & 1.03 & 0.59 & 4.53 & 4.60 & 0 \\
\multicolumn{1}{c|}{} & \multicolumn{1}{c|}{} & 64 & 3.88 & \textbf{0} & \textbf{0} & 0.85 & 0.46 & 5.37 & 6.15 & 0 \\
\multicolumn{1}{c|}{} & \multicolumn{1}{c|}{} & 128 & 4.50 & \textbf{0} & \textbf{0} & 0.63 & 0.31 & 6.33 & 10.25 & 0 \\
\multicolumn{1}{c|}{} & \multicolumn{1}{c|}{} & 256 & 4.96 & \textbf{0} & \textbf{0} & 0.39 & 0.18 & 8.88 & 22.33 & 0 \\ 
\bottomrule
\end{tabular}
}
\end{table*}

\begin{table*}[!t]
\centering
\caption{Quantitative comparison with the previous disambiguating methods on Discop using Baichuan2 and Chinese context.}\label{tab_exp}
\resizebox{0.9\textwidth}{!}{
\begin{tabular}{@{}cc|ccccccccc@{}}
\toprule
\multicolumn{2}{c|}{Method} & $k$ & Ave PPL & Ave KLD $\downarrow$ & Max KLD $\downarrow$ & \begin{tabular}[c]{@{}c@{}}Capacity $\uparrow$\\ (bits/token)\end{tabular} & Utilization $\uparrow$ & \begin{tabular}[c]{@{}c@{}}Total Time $\downarrow$\\ ($\times 10^{-2}$ seconds)\end{tabular} & \begin{tabular}[c]{@{}c@{}}Ave Time $\downarrow$\\ ($\times 10^{-6}$ seconds/bit)\end{tabular} & \begin{tabular}[c]{@{}c@{}}Total Error $\downarrow$\\ (\%)\end{tabular} \\ \midrule
\multicolumn{2}{c|}{\multirow{5}{*}{\begin{tabular}[c]{@{}c@{}}Discop\end{tabular}}} & 16 & 8.16 & 0 & 0 & 2.03 & 0.88 & 2.00 & 1.53 & 1.73 \\
\multicolumn{2}{c|}{} & 32 & 10.97 & 0 & 0 & 2.56 & 0.90 & 2.33 & 1.61 & 4.36 \\
\multicolumn{2}{c|}{} & 64 & 13.96 & 0 & 0 & 2.98 & 0.91 & 2.90 & 1.83 & 5.36 \\
\multicolumn{2}{c|}{} & 128 & 20.12 & 0 & 0 & 3.60 & 0.93 & 4.16 & 2.44 & 5.32 \\
\multicolumn{2}{c|}{} & 256 & 26.17 & 0 & 0 & 3.99 & 0.94 & 7.92 & 4.37 & 6.30 \\ \midrule

\multicolumn{1}{c|}{\multirow{15}{*}{\begin{tabular}[c]{@{}c@{}}Discop +\end{tabular}}} & \multirow{5}{*}{Basic} & 16 & 12.96 & 25.54 & 98.63 & 1.97 & 0.87 & 4.52 & 2.77 & 0 \\
\multicolumn{1}{l|}{} &  & 32 & 20.22 & 30.34 & 98.63 & 2.48 & 0.86 & 6.17 & 2.96 & 0 \\
\multicolumn{1}{l|}{} &  & 64 & 34.87 & 36.12 & 98.63 & 3.03 & 0.89 & 10.87 & 4.49 & 0 \\
\multicolumn{1}{l|}{} &  & 128 & 45.63 & 38.36 & 98.59 & 3.32 & 0.87 & 26.33 & 9.25 & 0 \\
\multicolumn{1}{l|}{} &  & 256 & 67.07 & 40.91 & 98.62 & 3.68 & 0.87 & 83.13 & 25.65 & 0 \\ \cmidrule(l){2-11} 

\multicolumn{1}{c|}{} & \multirow{5}{*}{MWIS} & 16 & 6.63 & 6.89 & 62.76 & 1.60 & 0.75 & 13.17 & 11.48 & 0 \\
\multicolumn{1}{l|}{} &  & 32 & 8.71 & 8.20 & 61.90 & 2.05 & 0.77 & 19.86 & 15.13 & 0 \\
\multicolumn{1}{l|}{} &  & 64 & 12.15 & 9.87 & 64.48 & 2.55 & 0.78 & 36.10 & 25.75 & 0 \\
\multicolumn{1}{l|}{} &  & 128 & 14.82 & 10.85 & 65.60 & 2.86 & 0.78 & 131.73 & 83.70 & 0 \\
\multicolumn{1}{l|}{} &  & 256 & 19.50 & 11.92 & 67.10 & 3.23 & 0.78 & 1048.35 & 671.89 & 0 \\ \cmidrule(l){2-11} 

\multicolumn{1}{c|}{} & \multirow{5}{*}{SyncPool} & 16 & 7.49 & \textbf{0} & \textbf{0} & 1.57 & 0.72 & 4.11 & 4.06 & 0 \\
\multicolumn{1}{l|}{} &  & 32 & 10.93 & \textbf{0} & \textbf{0} & 2.08 & 0.74 & 4.80 & 4.87 & 0 \\
\multicolumn{1}{l|}{} &  & 64 & 14.40 & \textbf{0} & \textbf{0} & 2.46 & 0.74 & 5.95 & 6.83 & 0 \\
\multicolumn{1}{l|}{} &  & 128 & 19.16 & \textbf{0} & \textbf{0} & 2.78 & 0.73 & 8.07 & 13.05 & 0 \\
\multicolumn{1}{l|}{} &  & 256 & 23.57 & \textbf{0} & \textbf{0} & 3.05 & 0.73 & 12.77 & 32.12 & 0 \\ 
\bottomrule
\end{tabular}
}
\end{table*}

\subsection{Setup}

In the experiments, we choose to conduct steganography experiments on English and Chinese, two of the most widely used languages worldwide. 
Meanwhile, unlike English, Chinese is a \textit{scriptio continua} without spaces between characters. Researchers believe that segmentation ambiguity has a more significant impact on {scriptio continua}~\cite{nozaki2022addressing}. 
In our experiments, we will also explore this by comparing the results of experiments conducted in English and Chinese.
We deploy the pretrained models \texttt{LLaMA2-7b}~\footnote{\url{https://huggingface.co/meta-llama/Llama-2-7b-hf}} and \texttt{Baichuan2-7b}~\footnote{\url{https://huggingface.co/baichuan-inc/Baichuan2-7B-Base}} from Hugging Face as the basic generative model for English and Chinese, respectively. Both tokenizers are implemented based on subwords.

To more intuitively assess how performance is affected by the number of elements in the candidate pool, we only use top-$k$ sampling to constrain the size of the initial candidate pool and the probability distribution. The reason for not changing the temperature is that it has no effect on the number of tokens. As for top-$p$, its effect is similar to that of top-$k$, but the size of the candidate pool may fluctuate significantly at each sampling step, which is not easy to be quantified intuitively. In the experiments, we set the truncation parameter $k=16,32,64,128,256$.
For each $k$, we select $100$ pieces of text from the IMDb dataset~\cite{maas2011learning}. We use the first three English sentences of each sample and their Chinese translations as the context for LLaMA2~\cite{touvron2023llama2} and Baichuan2~\cite{baichuan2023baichuan2} in different languages, and generate the subsequent 100 tokens conditioned on this context.

We compare our proposed disambiguating algorithm SyncPool with the basic~\cite{nozaki2022addressing} and MWIS-based~\cite{yan2023secure} token-removing disambiguating methods on a provably secure steganographic algorithm Discop~\cite{dingDiscopProvablySecure2023a}.

All experiments are carried out under the same hardware settings (CPU 3.00 GHz, 128 GB RAM, and NVIDIA RTX A6000).

\subsection{Metrics}

We assess the performance of disambiguation algorithms primarily based on three aspects: the effectiveness of disambiguation, its influence on the security of the steganographic method, and its impact on the efficiency of the steganographic method.
We evaluate the impact by comparing the steganographic performance before and after the integration of the disambiguation algorithm using metrics for assessing steganographic methods. 

\subsubsection{Effectiveness}

An effective disambiguation algorithm should reduce the error rate of the extracted secret message to $0\%$.
\begin{itemize}
    \item \textbf{Total Error:} the percentage of bits with decoding errors out of all embedded message bits.
\end{itemize}

\subsubsection{Security}

In generative linguistic steganography, our goal is to ensure that the stegotext, where the message is embedded, is indistinguishable from normally generated text where no message is embedded. 
We use KL divergence (KLD) to measure the distribution discrepancy between the original distribution provided by the language model during each token sampling period and the distribution used for steganographic sampling after disambiguation and steganographic methods. 
\begin{itemize}
    \item \textbf{Ave KLD:} the average KLD across all time steps.
    \item \textbf{Max KLD:} the maximum KLD across all time steps.
\end{itemize}
These two metrics indicate the average and maximum disruption of the original distribution by steganographic and disambiguation methods. Lower values are preferred.

We also use average perplexity to measure the perceptual imperceptibility of steganographic texts. 
\begin{itemize}
    \item \textbf{Ave PPL:} the average perplexity of generated texts.
\end{itemize}
The smaller the difference in average perplexity values between the steganographic texts generated before and after applying the disambiguation algorithm, the stronger the security of the disambiguation algorithm. Ave PPL can be calculated as:
\begin{equation}
    \textbf{Ave PPL} = \frac{1}{M}\sum_{j=1}^{M} 2^{-\frac{1}{N_j}\sum_{i=1}^{N_j}\log_{2}{\text{Pr}_j\left[x_i|x_{<i}\right]}},
\end{equation}
where $M$ is the number of text sequences tested, $N$ is the length of the sequence, $\text{Pr}_j\left[x_i|x_{<i}\right]$ is the probability of generating the $i$-th token given the previous tokens of the $j$-th text.

\subsubsection{Efficiency}

The efficiency of steganography encompasses the embedding rate and time consumption. We calculate the entropy utilization rate to assess the embedding rate of steganography. 
\begin{itemize}
    \item \textbf{Utilization:} the ratio of the total length of the embedded message to the sum of entropy across all time steps.
\end{itemize}
Additionally, we conduct timing experiments on language models for the ambiguity-unaware steganography generation process and the disambiguating steganography generation process. The time at which the model predicts the probability distribution is ruled out.
\begin{itemize}
    \item \textbf{Ave Time:} the average time consumed for embedding each bit of message, which is derived by dividing the total time of embedding by the total length of the embedded message.
\end{itemize}

\subsection{Results and Analysis}


We compared the effectiveness of our method with those of token-removal-based methods for resolving ambiguity. The English and Chinese experimental results are presented in~\autoref{tab_exp_en} and~\autoref{tab_exp}, respectively. 
In these tables, 
``Discop'' refers to the ambiguity-unaware provably secure steganographic algorithm that does not utilize any ambiguity resolution methods. 
``Basic'' and ``MWIS'' correspond to the basic version and the version with MWIS of the token-removal-based disambiguating algorithm, respectively. ``SyncPool'' refers to the method that we proposed in this paper.
The experimental results are described and analyzed as follows.

\subsubsection{Effectiveness}
\begin{table}[htbp]
\caption{
An example of a Discop stegotext generated by the Baichuan2 model, which encounters decoding errors. We use color blocks of different colors to represent different word segmentation. Tokens and messages that will result in errors during the extraction process are highlighted in red.}\label{tab_error}
\centering
\resizebox{\columnwidth}{!}{
\begin{tabular}{@{}l@{}}
\toprule

\begin{tabular}[c]{@{\hspace{10pt}}p{10cm}@{\hspace{10pt}}}\textbf{Secret Message:}\\ 
11111110 0011101100111001110111110011110100110010111100101011110 \dots 
\end{tabular} \\ \midrule

\begin{tabular}[c]{@{\hspace{10pt}}p{10cm}@{\hspace{10pt}}}\textbf{Stego Token IDs:}\\  
\dots 3203, 1376, \textcolor{blue}{7268, 15365}, 1738, 1773, 1376, 1352, 5099, 11198, 1377, 17620, 63499, 1360, 1374, 1352, 39663, 73, 7548 \dots
\dots
\end{tabular} \\ \midrule

\begin{tabular}[c]{@{\hspace{10pt}}p{10cm}@{\hspace{10pt}}}\textbf{Stegotext with Segementation Ambiguity:}\\ 
\textbf{The Sender:} \\

\dots women of\hlc[c1]{ Miss}\hlc[c2]{ouri} were some of the sexiest and attractive babes in the Midwest! Then \dots 
\end{tabular} \\ 

\begin{tabular}[c]{@{\hspace{10pt}}p{10cm}@{\hspace{10pt}}}
\textbf{The Receiver:} \\ 
\dots women of\hlc[c3]{ Missouri} were some of the sexiest and attractive babes in the Midwest! Then \dots 
\end{tabular} \\ \midrule

\begin{tabular}[c]{@{\hspace{10pt}}p{10cm}@{\hspace{10pt}}}\textbf{Retokenized Stego Token IDs:}\\ 
\dots 3203, 1376, \textcolor{red}{17858}, 1738, 1773, 1376, 1352, 5099, 11198, 1377, 17620, 63499, 1360, 1374, 1352, 39663, 73, 7548 \dots 
\end{tabular} \\ \midrule

\begin{tabular}[c]{@{\hspace{10pt}}p{10cm}@{\hspace{10pt}}}\textbf{Secret Message:}\\ 
11111110 \textcolor{red}{0101011101110101011011111101101011011110011111011010100} \dots 
\end{tabular} \\ \bottomrule

\end{tabular}}
\end{table}

Experiments reveal that regardless of whether the Llama2 model or the Baichuan2 model is used, a proportion of message bits will suffer decoding errors if disambiguating algorithms are not performed.
Overall, the total message decoding error rate is positively correlated with the number of candidate words. This trend is more pronounced when using the Baichuan2 model for steganography in Chinese.

~\autoref{tab_error} presents a stegotext generated by Discop using Llama2. It can be observed that when segmentation ambiguity leads to extraction errors, it not only affects the current token, but also results in changes to the probabilities corresponding to each subsequent token and the extractable message.
Therefore, in the case of errors, the impact of segmentation ambiguity is significant. It cannot be simply corrected by error correction codes.

All three disambiguating algorithms, including our method, are effective in reducing the total decoding error rate to zero.

\subsubsection{Security}
The KL divergence of the provably secure steganography method is zero, ensuring that the adversary does not gain any nongeligible advantage. However, existing ambiguity resolution methods disrupt the distribution that steganographic methods painstakingly maintain, causing a significant deviation between the distribution of the stegotext and the model's original distribution. This means that once these disambiguation methods are used, steganography is no longer provably secure. Our disambiguating method still maintains a KL divergence of zero, making it perfectly suitable for provably secure steganography.

In terms of average perplexity (Ave PPL), our proposed method maintains a similar level of perplexity to the original steganographic method.
The basic token-removal-based method significantly increased the perplexity of the generated stegotexts. 
The MWIS method yields stegotexts whose average perplexity is lower than that of the original steganography method, which does not mean that it is the best method. 
The calculated perplexity indicates the degree to which the output text conforms to the model's expectations. 
If a greedy sampling approach is used at each generation step, meaning always selecting the token with the highest predicted probability, the perplexity of the generated text will be the lowest. 
However, this does not align with the normal behavior of random sampling that an ordinary user would employ. 
Therefore, the standard for evaluating the average perplexity metric should be how closely it approaches the original ambiguity-unaware steganography method.
Under this criterion, our method performs the best, being the most difficult to distinguish from the original steganography method without disambiguation.

\subsubsection{Utilization Efficiency}

Our SyncPool method eliminates the amount of information lost in the distribution after detokenization and retokenization, reducing the total amount of entropy available for steganography. Therefore, a decrease in message embedding efficiency is inevitable.

As a comparison, the ambiguity-unaware Discop method achieves an entropy utilization rate ranging from $0.84$ to $0.88$ in the English context and from $0.88$ to $0.94$ in the Chinese context.
After disambiguation with our method, the entropy utilization rate of Discop is reduced.
In the Chinese context, our method's embedding capacity only decreased by less than 1 bit per token.
However, when using English contexts, the Discop method after disambiguation with SyncPool shows a noticeable decrease in the entropy utilization rate of the distribution predicted by Llama2.
Even so, when using a lower $k$ value such as $64$, we can embed an average of $0.85$ bits or more of information for each generated token. This indicates that the elimination of ambiguity does not affect the availability of provably secure steganographic methods.

Due to the alteration of the distribution, the entropy changes with each embedding, causing changes in embedding rates for the other two methods as well.

 

\subsubsection{Time Efficiency}

\begin{figure}[htbp]
    \subfloat[]{\includegraphics[width=0.49\columnwidth]{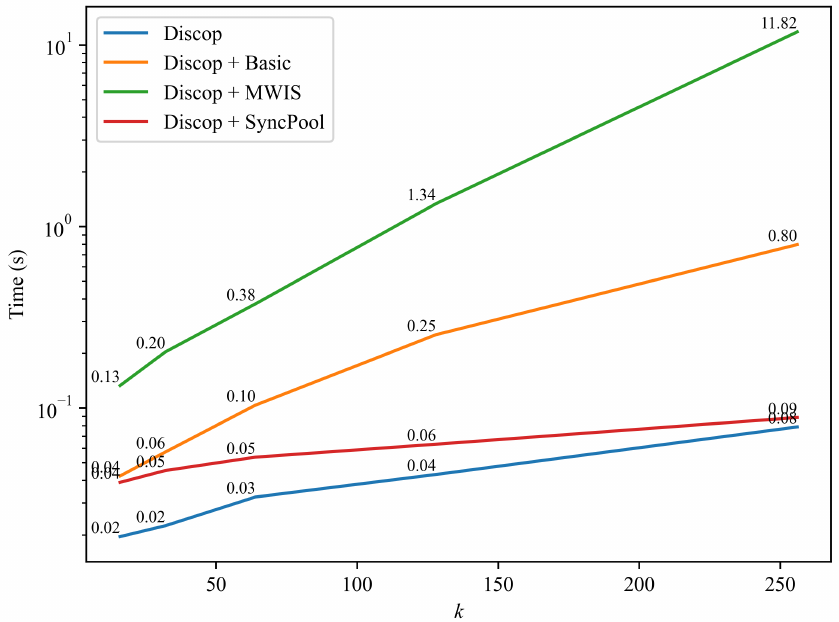}}
    \hspace{2pt}
    \subfloat[]{\includegraphics[width=0.49\columnwidth]{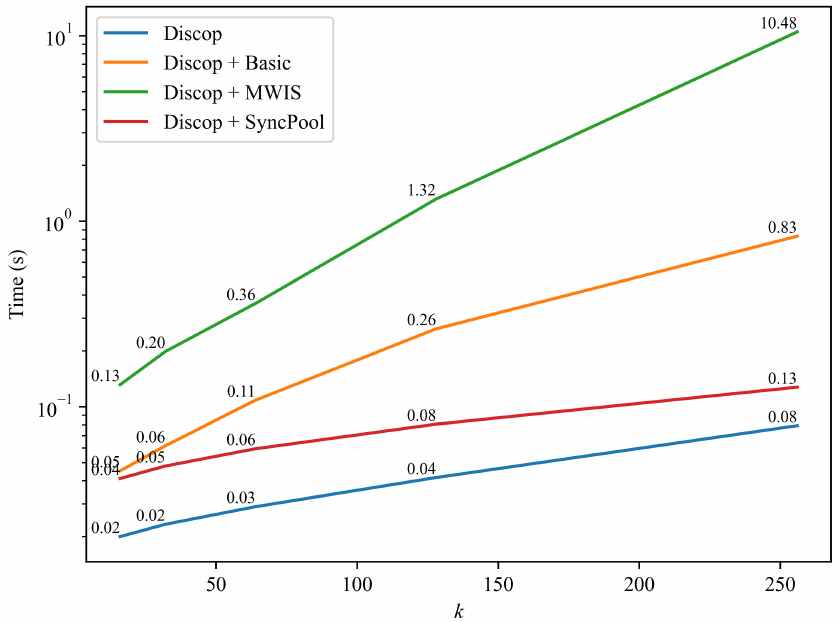}}
    \caption{Time consumption introduced by disambiguating algorithms. (a) Llama2; (b) Baichuan2.}
    \label{fig_time}
\end{figure}

\autoref{fig_time} shows the variation of embedding time (generation of $100$ tokens) with the value of $k$ for the steganographic embedding algorithm using different disambiguation methods.
Compared to ambiguity-unaware embedding, 
our disambiguating provably secure steganography does not consume much additional time.

\subsection{Ambiguity Frequency}

\begin{table}[t]
\caption{The Vocabulary Sizes and the Frequency of Prefix Relationships of Different Models.}\label{tab_ambiguity}
\centering
\resizebox{0.45\textwidth}{!}{
\begin{tabular}{@{}cc|c|ccccc@{}}
\toprule
\multicolumn{2}{c|}{\multirow{2}{*}{Model}} & \multirow{2}{*}{Vocal Size} & \multicolumn{5}{c}{Ambiguity Frequency} \\ 
\multicolumn{2}{c|}{} &  & 16 & 32 & 64 & 128 & 256 \\ \midrule
\multicolumn{1}{c|}{\multirow{2}{*}{English}} & \texttt{GPT-2} & 50,257 & 0.05 & 0.12 & 0.20 & 0.27 & 0.39 \\
\multicolumn{1}{c|}{} & \texttt{Llama-2-7b} & 32,000 & 0.16 & 0.26 & 0.41 & 0.57 & 0.78 \\ \midrule
\multicolumn{1}{c|}{\multirow{2}{*}{Chinese}} & \texttt{GPT-2-chinese} & 50,257 & 0.30 & 0.45 & 0.48 & 0.57 & 0.63 \\
\multicolumn{1}{c|}{} & \texttt{Baichuan2-7b} & 125,696 & 0.15 & 0.20 & 0.31 & 0.38 & 0.45 \\ \bottomrule
\end{tabular}
}
\end{table}

To further investigate the influence of segmentation ambiguity across various models, we conducted experiments on several common English and Chinese open-source models, \texttt{GPT-2}\footnote{\url{https://huggingface.co/openai-community/gpt2}}, \texttt{Llama-2-7b}, \texttt{GPT-2-chinese}\footnote{\url{https://huggingface.co/benjamin/gpt2-wechsel-chinese}}, and \texttt{Baichuan2-7b}. 
\autoref{tab_ambiguity} illustrates the vocabulary sizes of various models and the frequency of ambiguity under different top-$k$ truncation settings. Here, we randomly sample context $100$ times, with $100$ tokens sampled each time. The likelihood of segmentation ambiguity is represented by the proportion of tokens selected via random sampling in each step that share a prefix relationship with the other $k-1$ tokens. 
Overall, within the same language, as the number of tokens in the candidate pool increases, the frequency of ambiguity also tends to increase. Additionally, with smaller vocabulary sizes, the ambiguity frequency correspondingly increases. When comparing different languages, languages such as Chinese, which lack natural segmentation, have a greater probability of experiencing segmentation ambiguity than languages such as English. 



\subsection{Examples of Stegotext}

\begin{table}[t]
\centering
\caption{Examples of generating stegotext in different languages using Discop~\cite{dingDiscopProvablySecure2023a} with and without disambiguation on Baichuan2. $k=256$.}\label{tab_output}
\resizebox{\columnwidth}{!}{
\begin{tabular}{@{}l@{}}
\toprule
\begin{tabular}[c]{@{\hspace{10pt}}p{10cm}@{\hspace{10pt}}}\textbf{Context in English:}\\ When vapidness encouter doubtfulness, ideal may become a strange.\end{tabular} \\ \midrule
\begin{tabular}[c]{@{\hspace{10pt}}p{10cm}@{\hspace{10pt}}}\textbf{Discop~\cite{dingDiscopProvablySecure2023a}:}\\ Our daily habitual words and thoughts are basically unfeeling. It is like going into a house filled with a smell. In the meantime, this habitualness has to be shattered, as a new world with high ideal, which will bring us forward to an ideal world, is coming to view. To be born is to awakened.
People around the world are not indifferent to what kind of civilization they are now living. Yet I think they are all as much sleeping as to their \dots \end{tabular} \\ \midrule
\begin{tabular}[c]{@{\hspace{10pt}}p{10cm}@{\hspace{10pt}}}\textbf{Discop~\cite{dingDiscopProvablySecure2023a} + SyncPool:}\\ When one’s vapidness encounters fear, the ideal can be forgotten.
How to become a being again ? As for how and why the human being turns into the animal, since the very beginning of society, it was precisely a matter of this ? This issue is a truly difficult one, not only because it is difficult to grasp in its entirety, or even to get aware of the existence and the reality of this riddle; The only thing that can be ascertained with certainty \dots\end{tabular} \\ \midrule

\begin{tabular}[c]{@{\hspace{10pt}}p{10cm}@{\hspace{10pt}}}\textbf{Context in Chinese:}\\ \begin{CJK*}{UTF8}{gkai}无聊望见了犹豫，达到理想不太易。\end{CJK*}\end{tabular} \\ \midrule
\begin{tabular}[c]{@{\hspace{10pt}}p{10cm}@{\hspace{10pt}}}\textbf{Discop~\cite{dingDiscopProvablySecure2023a}:}\\ \begin{CJK*}{UTF8}{gkai}不必灰心，不必愁，一切归於零。希望重燃一起望，困难永驱驰。最羡是情花一簇,无艳亦艳丽，开遍遍野，开遍满地。最羡是情花一簇，带给我一双手,一对臂膀,更抱紧了我俩。永远在爱里，花儿总这样开下去。鲜花铺满了整个大草原，将歌声传遍世界每个角落，愿人人都来唱歌，让全世界 \dots\end{CJK*}\end{tabular} \\ \midrule
\begin{tabular}[c]{@{\hspace{10pt}}p{10cm}@{\hspace{10pt}}}\textbf{Discop~\cite{dingDiscopProvablySecure2023a} + SyncPool:}\\ \begin{CJK*}{UTF8}{gkai}如果你不是这样，那你就没有希望了。如果你是这样的，那麽，不要使你任性的个性去违反你的理想，就应当使你不屈服于理想，而使你趋向於理想。 但是，这样你又不能使你抱有的理想成为成功的把握，因为成功需要另外许多的条件。不能使你成功的理想，才是符合你个性的理想；不能使你存在的个性，不是你的理想，因为你在社会上存在，所以这种理想，你得去实现它。你先得有一种使你存在的个性 \dots\end{CJK*}\end{tabular} \\ \bottomrule
\end{tabular}}
\end{table}

\autoref{tab_output} provides an example of employing Discop with and without our proposed disambiguating algorithms on Baichuan2 to continue writing 100 tokens under $k=256$ truncation. The semantics of stegotexts are all relevant to the given context.
\section{Discussion}


Our disambiguating approach has a considerable limitation, that is, we have reduced to some extent the entropy utilization of the provably secure steganography method with respect to the model's original distribution. 

Provably secure steganography aims to perfectly utilize the distribution predicted by the model, achieving an entropy utilization rate of $1.0$. 
From the perspective of these steganography methods, each token in the model candidate pool is perceived as unique, and based on its distribution, it can represent different messages. 
However, when we consider segmentation ambiguity, two groups of different tokens with the same prefix (such as ``\_any" + ``thing" and ``\_anything") are completely identical after detokenization, resulting in loss of the information they can express.
We use a shared CSPSNR for synchronization between both parties to assist the receiver in determining the current token and completing extraction during the autoregressive process. However, this information has indeed been lost. Therefore, the embedding rate of steganography inevitably decreases. 
The extent of reduction is related to the likelihood of encountering segmentation ambiguity and the distribution of ambiguity pools after grouping.

It is important to emphasize that ambiguity frequency does not necessarily lead to decoding errors. A comparison of the data in~\autoref{tab_exp_en} and~\autoref{tab_exp} reveals that although the English Llama2 model has a higher frequency of ambiguity issues, the Baichuan2 model in the Chinese context experiences more decoding errors.
To ensure accurate extraction of the message in every steganographic process, it is necessary to address all possible instances of ambiguity.
Therefore, although the number of decoding errors in English experiments is fewer than in Chinese experiments, the cost of embedding rate for eliminating ambiguity is higher.



\section{Conclusion}

In this paper, we analyze the segmentation ambiguity problem that provably secure linguistic steganography must address to become practical and explain why existing disambiguating algorithms are inadequate for provably secure steganography.
We propose a novel, effective, secure disambiguating method based on ambiguity pool grouping and synchronous sampling, which does not alter the original candidate pool or probability distribution, thus enabling integration with provably secure steganography methods.
Our method utilizes a shared PSNR to perform synchronized sampling in ambiguity pools, ensuring that the receiver obtains a uniquely determined decoding result.
We conduct experiments on a provably secure steganography method, and the results demonstrate that our method preserves the distribution, incurs no significant decrease in utilization efficiency, and does not compromise the security of the steganography method.

\ifCLASSOPTIONcaptionsoff
  \newpage
\fi



\bibliographystyle{IEEEtran}
\bibliography{main}
\end{document}